# Conductance Quantization in Schottky-gated Si/SiGe Quantum Point Contacts

G. Scappucci,[1,*] L. Di Gaspare,[1] E. Giovine,[2] A. Notargiacomo,[2] R. Leoni,[2] and F. Evangelisti[1,2]

[1]*Dipartimento di Fisica "E. Amaldi", Università Roma TRE, V. Vasca Navale 84, 00146 Roma, Italy.*
[2]*Istituto di Fotonica e Nanotecnologie, IFN-CNR, Via Cineto Romano 42, 00156 Roma, Italy.*

**Abstract.** We report on the fabrication and electronic transport characterisation of Schottky-gated strongly confined Si/SiGe quantum point contacts (QPC). At zero magnetic field and T=450mK the QPC conductance as a function of gate voltage shows a quantization in units of $e^2/h$, indicative of transport through 1D modes which appear to lack both spin and valley degeneracy.

## INTRODUCTION

Despite the high quality reached in the last decade for the Si/SiGe two-dimensional electron gas (2DEG), only recently significant progress has been achieved in obtaining high confinement of charge carriers and effective gating action on Si/SiGe quantum devices.[1, 2]
This allows also for the Si electron gas conductance investigations that have previously been restricted to GaAs systems. In this paper we report on the fabrication of highly confined Si/SiGe etched quantum point contact efficiently controlled by Schottky gates and discuss the presence of a surprising $e^2/h$ conductance quantization.

## EXPERIMENTAL

We fabricated Schottky-gated QPCs on high mobility Si/SiGe 2DEG. The Si/SiGe 2DEG heterostructure was grown by chemical vapour deposition on Si(001) substrates. At T=300mK, the value of the 2DEG carrier density, estimated from low-field Hall measurements on mesa-etched Hall bars, was $9.8 \times 10^{11} \text{cm}^{-2}$ and the electron mobility $4.1 \times 10^4 \text{cm}^2/\text{Vs}$. From these values we estimate a mean-free path of the 2D unconstrained carriers of the order of 500 nm. [2] The QPCs were obtained by confining the 2DEG in a double-bend like geometry by electron-beam lithography (EBL) and reactive ion etching. In Fig. 1 we report a scanning electron micrograph of a complete device.

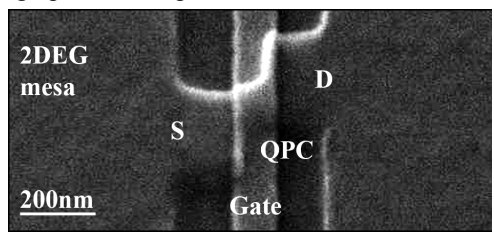

**FIGURE 1.** Scanning electron micrograph of a Si/SiGe etched quantum point contact controlled by a Schottky top-gate.

The QPC is formed by the narrow conducting channel (lithographic width $w \sim 170$ nm, effective width much smaller due to lateral depletion) that originates at the 2DEG mesa protrusions labeled *S* and *D* in Fig. 1. By EBL and lift-off we patterned a 5/30-nm-thick Ti/Au 100nm-wide finger gate crossing the central constriction and folding along the etched surfaces of the constriction; the electric field imposed by the gate on the lateral walls is screened by the surface states so that in this configuration the gate varies the carrier concentration without changing the width of the QPC.[3] At T=450mK the leakage from the Schottky gate (active area less than $0.1 \times 0.16 \mu m^2$) to the 2DEG was negligible (leakage current $I_{GATE} < 0.2$ pA) in the -2V to +1V gate voltage range. This large available working range enabled a full control of electronic transport through the QPC from conduction to depletion.

## RESULTS AND DISCUSSION

Electronic transport characterisation of the QPC devices was performed at T = 450 mK in a $^3$He refrigerator at zero magnetic field using standard ac low frequency lock-in techniques. The linear conductance G versus the gate voltage $V_G$ is reported in Fig. 2(a). The curve was corrected to take in account for a series resistance $R_S$= 19.4 k$\Omega$, originating from both the outer 2DEG mesa structure as well as the source and drain contacts. The conductance shows clearly a staircase with plateau-like structures close to multiple integers of $e^2/h$, suggesting transport through 1D modes in with both spin and valley degeneracy appear removed at zero magnetic field. The data were highly reproducible upon sweeping the gate from positive to negative voltages or cycling the temperature from 450 mK to room temperature.

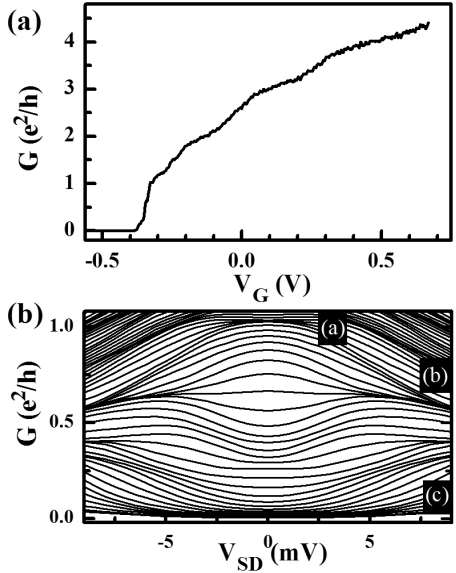

**FIGURE 2.** Electronic transport characterisation of the Si/SiGe QPC at T = 450 mK. (a) Conductance vs. gate voltage exhibiting $e^2/h$ quantization. (b) Finite source-drain bias spectroscopy showing the presence of a integer plateau that at finite $V_{SD}$ evolves in a half $e^2/h$ plateau.

To gain insight into the $e^2/h$ quantization, finite source-drain bias spectroscopy was performed by measuring the non-linear conductance G versus source-drain bias $V_{DS}$ at different $V_G$. In Fig. 2(b) we report the curves measured by changing the gate voltage in the range where the linear conductance in Fig. 2(a) develops the $e^2/h$ feature. The data were corrected to take in account the presence of an electrostatic self-gating effect and, again, of a series resistance. The G-$V_{SD}$ curves show an evolution with the gate voltage as expected for the $e^2/h$ quantization. Several curves bundle at $V_{SD} \sim$ 0 mV at the conductance value $e^2/h$ in the gate voltage range where the G-Vg curve developed the $e^2/h$ plateau [$V_G \sim$ -0.3125 V, label (a)]; this integer plateau evolves into a half-integer plateau (G $\sim$ 0.5 $e^2/h$) as the source-drain bias is increased to a finite value [$V_{SD}$~4mV, label (b)]. From the $V_{SD}$ position of the half-integer plateau we estimate the energy spacing between the first two 1D subband $\Delta E \sim$ 4 meV, indicating that significant quantum confinement is achieved in the device. At large negative gate values [label (c)], there is no conduction at $V_{SD} \sim$ 0 mV meaning that the Fermi level is below the onset of the first 1D band.

The behaviour of the linear and non-linear conductance upon changing the gate bias can be explained considering adiabatic transmission through 1D modes with complete removal of valley and spin degeneracies. In a Si 1D electron gas the conductance quantization was previously found in units of $4e^2/h$, as expected in a valley and spin degenerate electron gas.[4] In our devices the removal of the valley degeneracy is likely to be the result of the strong confining potential; more unexpected is the presence of the $e^2/h$ plateau at zero magnetic field and evidence for the $e^2/h$ quantization. Recently an unexpected $e^2/h$ quantization was reported also in carbon nanotubes.[5] These findings might be closely related to the "0.7 anomaly", a spin-related phenomenon observed at zero magnetic field in clean 1D GaAs systems which is assumed to signal the occurrence of non negligible correlation effects. [6]

## ACKNOWLEDGMENTS

This work was partially supported by the FIRB project RBNE01FSWY "Nanoelettronica" and the FISR project "Nanotecnologie per dispositivi di memoria ad altissima densità".

## REFERENCES

\* Present address: School of Physics, University of New South Wales, Sydney NSW 2052, Australia.
1. G. Scappucci, L. Di Gaspare, F. Evangelisti, E. Giovine, A. Notargiacomo, R. Leoni, V. Piazza, P. Pingue, and F. Beltram, *Phys. Rev. B* **71**, 245311 (2005).
2. G. Scappucci, L. Di Gaspare, E. Giovine, A. Notargiacomo, R. Leoni , and F. Evangelisti, *Phys. Rev. B*, in press (2006).
3. G. Curatola and G. Iannaccone, *J. Appl. Phys*. **95**, 1251 (2004).
4. D. Tobben, D. A. Wharam, G. Abstreiter, J. P. Kolthaus, and F. Schaffler, *Semicond. Sci. Technol.* 10, 711 (1995).
5. M. J. Biercuk, N. Mason, J. Martin, A. Yacoby, and C. M. Marcus, *Phys. Rev. Lett.* **94**, 026801 (2005).
6. K. J. Thomas, J. T. Nicholls, M. Y. Simmons, M. Pepper, D. R. Mace, and D. A. Ritchie, *Phys. Rev. Lett.* **77**, 135 (1996).